\documentclass{PoS}

\usepackage{physics}
\usepackage{amsmath}
\usepackage{siunitx}
\usepackage{mathrsfs}
\usepackage{amssymb}
\usepackage{bbm}
\usepackage{mathtools}
\usepackage{dsfont}
\usepackage[colorinlistoftodos]{todonotes}

\usepackage{collref}

\title{Confronting Scherk-Schwarz orbifold models with LHC data}
\ShortTitle{Scherk-Schwarz models \& LHC data}
\author{\speaker{Dumitru Dan Smaranda}\\
        School of Physics and Astronomy - University of Glasgow, Glasgow, UK \\
        E-mail: \email{d.smaranda.1@research.gla.ac.uk}}
\author{David J Miller \\
       School of Physics and Astronomy - University of Glasgow, Glasgow, UK
       \\ E-mail: \email{david.j.miller@glasgow.ac.uk} }

\abstract{ We will outline our recent efforts aimed at analysing a class of models known as orbifold GUTs and their phenomenology in a variety of minimal and non-minimal settings.  We examine the minimal $SU(5)$ models, rule them out, and proceed by extending them with an additional scalar field along with a gauge extension via $SU(5)\times U(1)$ models. We end up by commenting on the future improvements needed to more accurately handle exclusions along with tracing the $U(1)$ gauge extensions to more complete $6D$ theories.}

\FullConference{The 39th International Conference on High Energy Physics (ICHEP2018)\\
		4-11 July, 2018\\
		Seoul, Korea}

\begin{document}


\section{Introduction}

After the Higgs discovery, supersymmetry (SUSY) has had to face significant exclusions from the LHC data.
Indeed the constrained Minimal Supersymmetric Standard Model, which gives all the supersymmetry breaking parameters a common value at a high scale, is almost completely ruled out~\cite{Aad:2015iea}. However, typical supersymmetric models are complicated by having over 100 additional free parameters, so plenty of the parameter space remains to be explored. To this extent non-minimal supersymmetric models fit within these regions providing a continuing motivation for the LHC to search for them.

In this presentation we'll outline our efforts on how Scherk-Schwarz (SS) compactifications \cite{Scherk1,Scherk2}  affect a variety of extra dimensional GUT models with Kaluza Klein modes \cite{Kaluza:1921tu,Klein:1926tv}, and see if they agree with phenomenological constraints imposed by electroweak symmetry breaking and low energy experiments. We start of by exploring the basic model proposed in \cite{Barbieri1, Barbieri2}, and then move on to study scalar and $U(1)$ gauge extensions.


\section{Theory and Models}

Throughout this paper we'll be working on a $5D$ compactified space $\mathcal{M}_4 \times S^1/\mathds{Z}_2$ with $\mathcal{N}=1$ SUSY. The SS action that we employ breaks the $5D, \mathcal{N}=1 $ supersymmetry  to $ 4D, \mathcal{N} = 1$ on the brane at $y=0$, the Higgs flavour symmetry $SU(2)_H \rightarrow U(1)_H$ and the gauge symmetry $SU(5) \rightarrow G_{SM}$ (note that we use the same gauge breaking for the $U(1)$ extension).
The full form of $Z, T$ is the one used in \cite{Barbieri1, Barbieri2}.

Using these will in turn provide a soft SUSY breaking Lagrangain, which will depend on the fermionic  matter placement (i.e. brane or bulk).

Since the basic model in \cite{Barbieri1, Barbieri2} will fail to produce the right Higgs mass we'll move on and extend the Higgs sector via a  scalar extension: $W = \lambda H_u H_d S + \frac{1}{3} \kappa S^3 $, which will produce soft SUSY breaking masses from the SS action depending on the scalar placement (brane or bulk).


\section{Methodology and Constraints}

High scale parameters are introduced at the GUT scale and are run down to low energies using the FlexibleSUSY [v.2.0.1] \cite{FlexibleSUSY} spectrum generator with two-loop Renormalisation Group Equations (RGEs), to produce electroweak symmetry breaking and a low energy spectrum. FlexibleSUSY relies on SARAH [v.4.12.2] \cite{SARAH} to generate the RGEs and the tadpole equations.

We check our model against LHC bounds and constraints from the ATLAS and CMS collaborations \cite{ATLASSUSY, ATLASEXOTICS, CMSSUSY}, which in our case comprise of:
 a Higgs mass between  $123 \leq m_H \leq 127 $ \SI{}{GeV} (where we've assumed a \SI{2}{GeV} theoretical uncertainty arising from FlexibleSUSY)\cite{HiggsATLAS,HiggsCMS}
 ;  a gluino  mass larger than $m_{ \tilde{g} } \geq 2$ \SI{}{TeV} \cite{Aaboud:2017vwy, Sirunyan:2017pjw}; a neutralino mass larger than  $m_{ \tilde{\chi}_1^0 } \geq 537$ for $\tan\beta \in [10, 50]$ \cite{ATLAS:2014fka};  a stop mass larger than   $ m_{ \tilde{t} } \geq \SI{1}{TeV}$ \cite{Aaboud:2017aeu} ;  a chargino mass larger than  $m_{\tilde{\chi}^\pm_1 } \geq 460$ \SI{}{GeV} \cite{CharginoSUSY}; an extra $Z'$ gauge boson with a mass larger than  $m_{Z'} \geq 2.4$ \SI{}{TeV} \cite{Aaboud:2017sjh} for the $U(1)$ extensions.

To check the dark matter relic density we use MicrOmegas \cite{MicrOmegas}. The dark matter relic density bound is in accordance with the latest Planck data \cite{PlanckData} :
 $ \Omega_c h^2 = 0.1157 \pm 0.0023$,  where we consider a $10\%$ uncertainty from the mass difference from MicrOmegas and FlexibleSUSY, and accept all points with a dark matter relic density smaller than $\Omega_c h^2 = 0.1275$.


\section{Results and Conclusions}

Throughout our scans we found that the basic $SU(5)$ model proposed in \cite{Barbieri1, Barbieri2} cannot produce the appropriate Higgs mass with brane or bulk matter. We then moved on to trying a scalar extension of the $SU(5)$ model which yielded in a better result concerning the Higgs, but failed in the end to meet the SS constraint. This will be further explored in future work to try and more accurately quantify the SS uncertainty resulting from threshold corrections. Furthermore to try and bypass the SS constraint we looked at a scalar extension with a trivialised $SU(2)_H$ symmetry. In this case we  got the right Higgs mass but the model was eliminated due to LHC cuts and dark matter relic density constraints.

\begin{figure}[h!]
 \includegraphics[scale = 0.265]{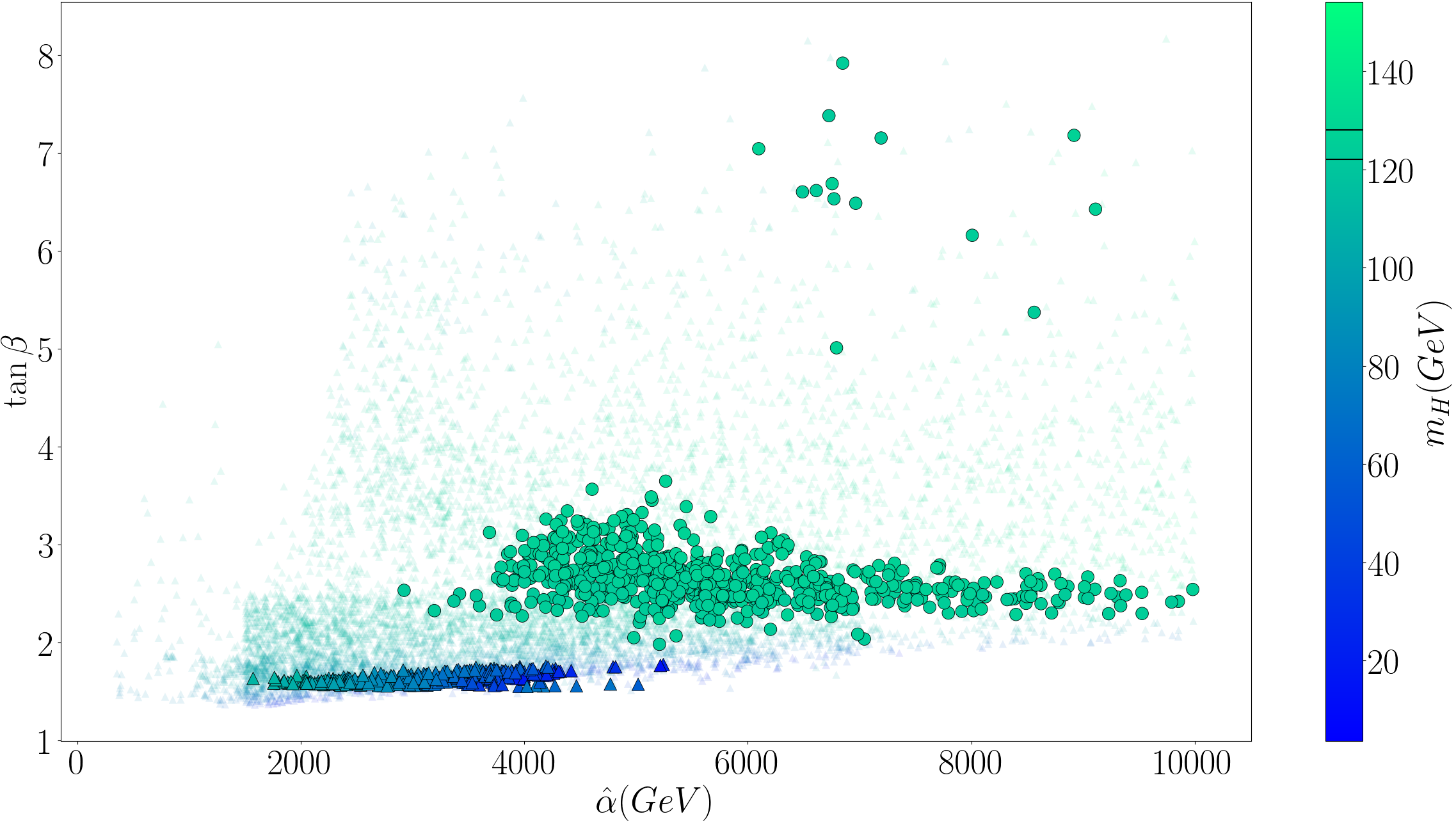}
 \caption{$SU(5)\times U(1)_N$ plus bulk Scalar $\mathscr{S}$ with brane matter $T,F$. The points marked by triangles represent points originating from SS breaking, and the points marked by circles  represent points with the right Higgs mass. The transparent points produce EWSB but do not pass LHC constraints. }\label{UMSSM_BuS_BrMat}
\end{figure}

Finally the additional scalar within the $U(1)_N$ extension framework resulted in a similar scenario where the points that obeyed the SS constraint did not produce an appropriate Higgs mass (see Figure \ref{UMSSM_BuS_BrMat}).  This has in turn prompted future work in which we will treat the spectrum in Figure \ref{UMSSM_BuS_BrMat} as a remnant of a $6D, E_6$ theory (e.g. as in \cite{Nevzorov}) which would  arise from a more complicated accidental flavour symmetry induced by the different $27, \overline{27}$ representations delivering ``a different SS constraint''.

To summarise, the basic model cannot produce the right Higgs mass, the naive scalar extensions don't quite work and we are looking on quantifying threshold effects along with exploring more complicated theories that can accommodate SS breaking.

\bibliographystyle{JHEP}
\bibliography{myBibliography}

\end{document}